\documentclass[english,aps,twocolumn]{revtex4}
\usepackage[T1]{fontenc}
\usepackage[latin9]{inputenc}
\usepackage{amstext}
\usepackage{graphicx}
\usepackage{amssymb}
\usepackage{esint}

\makeatletter
\@ifundefined{textcolor}{}
{%
 \definecolor{BLACK}{gray}{0}
 \definecolor{WHITE}{gray}{1}
 \definecolor{RED}{rgb}{1,0,0}
 \definecolor{GREEN}{rgb}{0,1,0}
 \definecolor{BLUE}{rgb}{0,0,1}
 \definecolor{CYAN}{cmyk}{1,0,0,0}
 \definecolor{MAGENTA}{cmyk}{0,1,0,0}
 \definecolor{YELLOW}{cmyk}{0,0,1,0}
 }

\makeatother

\usepackage{babel}

\begin{document}

\title{Octupole deformation properties of the Barcelona Catania Paris energy
density functionals}

\author{L.M. Robledo}

\email{luis.robledo@uam.es}

\affiliation{Dep. F\'{\i}sica Te\'orica (M\'odulo 15), Universidad Aut\'onoma de Madrid,
28049 Madrid, Spain}

\author{M. Baldo}

\affiliation{Instituto Nazionale di Fisica Nucleare, Sezione di Catania, Via Santa
Sofia 64, I-95123 Catania, Italy}

\author{P. Schuck}

\affiliation{Institut de Physique Nucl\'eaire, CNRS, UMR8608, F-91406 Orsay, France}

\author{X. Vi\~nas}

\affiliation{Departament d'Estructura i Constituents de la Mat\`eria and Institut
de Ci\`encies del Cosmos, Facultat de F\'{\i}sica, Universitat de Barcelona,
Diagonal \emph{647}, 08028 Barcelona, Spain}

\begin{abstract}
We discuss the octupole deformation properties of the recently proposed
Barcelona-Catania-Paris (BCP) energy density functionals for two set
of isotopes, those of radium and barium, where it is believed that
octupole deformation plays a role in the description of the ground
state. The analysis is carried out in the mean field framework (Hartree-
Fock- Bogoliubov approximation) by using the axially symmetric octupole
moment as a constraint. The main ingredients entering the octupole
collective Hamiltonian are evaluated and the lowest lying octupole
eigenstates are obtained. In this way we restore, in an approximate
way, the parity symmetry spontaneously broken by the mean field and
also incorporate octupole fluctuations around the ground state solution.
For each isotope the energy of the lowest lying $1^{-}$state and
the $B(E1)$ and $B(E3)$ transition probabilities have been computed
and compared to both the experimental data and the results obtained
in the same framework with the Gogny D1S interaction, which are used
here as a well established benchmark. Finally, the octupolarity of
the configurations involved in the way down to fission of $^{240}$Pu,
which is strongly connected to the asymmetric fragment mass distribution,
is studied. We confirm with this thorough study the suitability of
the BCP functionals to describe octupole related phenomena.
\end{abstract}
\maketitle

\section{Introduction}

The ground state and low lying excited states of many atomic nuclei
all over the nuclide chart show quadrupole deformation in their intrinsic
states \cite{Bohr.75}. This property has profound consequences in
the low lying spectrum of those nuclei as well as in their decay patterns
\cite{Ring.80,Bender.03}. Octupole deformation is not as common
as quadrupole deformation as a characteristic of the ground state
of atomic nuclei, but its consequences are important for understanding
nuclear properties of several actinide nuclei around radium and several
rare earth around barium \cite{Butler.96}. The octupole operator
has negative parity and therefore a non-zero octupole deformation
means that the intrinsic state has lost reflection symmetry and acquired
a pear-like shape. The quantum interference between the two degenerate
intrinsic states with pear-shaped matter distributions pointing upward
and downward (i.e. with the same absolute value of the octupole moment
but opposite sign) restores the parity quantum number and leads to
the presence in the spectrum of a doublet with opposite parities \cite{Bohr.75,Butler.96}.
The energy splitting between the two members of the doublet strongly
depends upon the properties of the barrier separating the two degenerate
intrinsic states with opposite octupole moment. In deformed even-even
nuclei it is possible that the negative parity member of the ground
state multiplet , the lowest lying $1^{-}$ state, can be located
below the lowest $2^{+}$state leading to the appearance of alternating
parity rotational bands, which are clear signatures of octupole deformation.
Also, the two members of the doublet will be connected by strong $B(E1,1^{-}\rightarrow0^{+})$
transition probabilities from the $1^{-}$ to the ground state. The
next member of the negative parity rotational band is a $3^{-}$ that
rapidly decays to the $0^{+}$ground state by means of strong $B(E3,3^{-}\rightarrow0^{+})$
transition probabilities. Although there are several known examples
of alternating parity rotational bands at low spins, the alternating
behavior usually appears at high spins as a consequence of the stabilizing
effect of angular momentum on the octupolarity of the system: moments
of inertia increase with the octupole moment and therefore configurations
with higher octupole moments are the more lowered by increasing angular
momentum. 

The appearance of octupole effects is strongly linked to the position
of the Fermi energy in the single particle spectrum of the underlying
mean field \cite{Bohr.75,Butler.96}. The reason is that octupolarity
is enhanced when, in a given major shell, the intruder orbital interacts
(via a particle-hole excitation) with a nearby normal parity orbital
with three units less of angular momentum ($3\hbar$ is the amount
of angular momentum carried out by the octupole operator). This happens
between the $j_{15/2}$ and $g_{9/2}$ spherical orbitals, the $i_{13/2}$
and $f_{7/2}$ and the $h_{11/2}$ and $d_{5/2}$. Those regions
where both protons and neutrons feel a strong octupole interaction
is where octupole related effects are expected to be more pronounced.
For example, in the region around $_{88}^{224}$Ra, the Fermi level
of protons is located around the $f_{7/2}$orbital that can then interact
strongly with the empty $i_{13/2}$. Besides, Fermi level of neutrons
is around the $g_{9/2}$ orbital, which strongly interacts through
the octupole interaction with the $j_{15/2}$ one. Similar arguments
apply to the region around $^{144}$Ba with the orbitals $d_{5/2}$
and $h_{11/2}$ for protons and the $f_{7/2}$ and $i_{13/2}$ for
neutrons responsible for octupolar effects.

Last but not least, octupolarity also plays a relevant role in the
asymmetric fission decay mode because an octupole deformed pathway to fission
naturally explain the observed asymmetric mass fragment distribution
of several actinide parent nuclei \cite{Fission}.

In this article we want to check the ability of the recently proposed
Barcelona-Catania-Paris (BCP) energy density functionals \cite{Baldo.08},
in dealing with the octupole degree of freedom in finite nuclei. We 
will compare the results provided by the BCP functionals
in some test cases with experimental values, when available, and with
the results obtained using the Gogny D1S interaction that we take
here as a benchmark. The BCP energy density functionals consist of
a bulk part which is fully microscopic and comes from the nuclear
and neutron equation of state \cite{Baldo.04} which are parametrized
in a polynomial form complemented by additional terms accounting for
finite size effects (see \cite{Fayans.98,Fayans.01} for functionals
inspired in the same principles). In addition to the Coulomb term and the spin-orbit
contribution, which is taken exactly as in the Skyrme or Gogny forces,
we add a purely phenomenological finite range term for describing
properly the nuclear surface. To deal with open-shell nuclei we still
include in the BCP functionals a zero range density-dependent pairing
interaction fitted to reproduce the nuclear matter gaps obtained with
the Gogny force \cite{Garrido.99}. The only free parameters of these
functionals are the isospin like (L) and unlike (U) strengths of the surface
term, the range of the Gaussian form factor used to give a finite
range to the surface term and the strength of the spin-orbit interaction
\cite{Baldo.08}. These free parameters are adjusted in the usual
way to reproduce the ground state energy and charge radii of some
selected spherical nuclei. With these ingredients, the BCP functionals
give an excellent description of 161 even-even spherical nuclei with
rms values for the ground-state energies (1.77 MeV and 2.06 MeV) for 
BCP1 and BCP2, respectively) and charge radii that are 
comparable to the ones obtained with well reputed interactions/functionals
like Skyrme SLy4 (1.71 MeV), 
Gogny D1S (2.41 MeV) or the  Relativistic  NL3 parametrization (3.58 MeV).
Apart from the advantages already mentioned in \cite{Baldo.08},
the BCP functionals are advantageous in its application to finite
nuclei because of its reduced computational cost as compared
to Gogny (BCP is a factor between 6 to 10 faster) or even Skyrme
(comparable computational cost). 
Also the appearance of integer powers of the density in the bulk 
part of the functional, which is a consequence
of the specific fit to the nuclear matter results, makes much easier
to deal with the self-energy problem that plagues beyond mean field
calculation \cite{Duguet.09}. Using these BCP functionals we have
also explored quadrupole deformation properties \cite{robledo.08}.
We find a behavior similar to that obtained using the Gogny D1S force
widely used to this end. This fact give us confidence in using the
BCP functionals to study nuclear properties related to deformation.
As the BCP functionals are aimed at describing not only masses and radii
but also the low lying spectrum over all the nuclide chart, it is
necessary to check whether the very reasonable results regarding quadrupole
collectivity can also be extended to the octupole deformation case.
To check that this is the case, we have carried out mean field Hartree-
Fock- Bogoliubov (HFB) calculations with the BCP energy density functional
as well as the Gogny \cite{Decharge.80} D1S \cite{Berger.84} interaction
to test the response of the system to the octupole degree of freedom.
To be more precise, we have used a constraint in the axially symmetric
octupole moment to generate potential energy curves (PECs) to search
for  octupole deformed minima as well as to study the stiffness of
those (and other) minima against changes in the octupole degree of
freedom. These PECs are computed for several isotopes of radium from
$^{216}$Ra to $^{232}$Ra and of barium from $^{140}$Ba until $^{150}$Ba.
In addition to the PEC, the calculation of the corresponding collective
inertias allows the evaluation of the $1^{-}$ excitation energy as
well as $B(E1)$ and $B(E3)$ transition probabilities in the framework
of the collective Schr\"odinger equation (CSE) method. The results will
be compared with experimental data, when available, as well as with
the results obtained with the Gogny D1S force. It should be mentioned
that the Gogny D1S results have already been reported in Refs. \cite{Robledo.87,Egido.89,Egido.90}
and similar calculations with the Skyrme interactions exist in the
literature \cite{Bonche.86}. Finally, the octupole properties of
the fission valley of $^{240}$Pu will also be briefly discussed and
compared to those of Gogny D1S.

\section{Theoretical tools}

To solve the Hartree-Fock-Bogoliubov (HFB) equation \cite{Ring.80}
the quasi-particle operators of the Bogoliubov transformation have
been expanded in a harmonic oscillator (HO) basis big enough as to
warrant convergence of the results with the basis size. The expansion
coefficients have been determined by means of the gradient method
which relies on the parametrization of the mean field (HFB) energy
in terms of the parameters of the Thouless expansion of the most general
HFB wave functions. Within the gradient method, the HFB problem is
recast in terms of a minimization (variational) process of the mean
field energy and the search of the minimum is performed by following
the direction of the gradient in the multidimensional space of parameters.
The advantage of this method over the more traditional one of successive
diagonalizations is in the way the constraints are implemented, which
allows a larger number of them to be treated at once. Axial symmetry
has been preserved in the calculation implying the use of an axially
symmetric HO basis made up of the tensor product of two dimensional
HO wave functions times one dimensional HO ones. Along with the octupole
moment constraint associated to the multipole operator $\hat{Q}_{3}=r^{3}Y_{30}$
and used to generate the PEC's, we have included a constraint on the
center of mass of the nucleus (i.e. the mean value of $r^{1}Y_{10}$
has been set to zero) to prevent spuriousness associated to the center
of mass motion, to slip into the results. As a consequence of the
axial symmetry imposed in the HFB wave functions, the mean values
of the $Q_{\lambda\mu}$ multipole operators with $\mu\ne0$ are zero
by construction. 

The information given by mean field theories is restricted to the
energy and shape of the -generally- deformed ground state. To restore
the parity symmetry broken by the mean field approximation and 
to describe the dynamics of the collective excited states it
is mandatory to go beyond the mean field approximation. With this
in mind, the octupole degree of freedom $Q_{3}=\langle\psi|\hat{Q}_{3}|\psi\rangle$
(where $|\psi\rangle$ is the HFB intrinsic wave function) has been
used to build up a collective Hamiltonian based on the Generator Coordinate
Method (GCM) and the Gaussian Overlap Approximation (GOA) \cite{Brink.68,Giraud.74,Reinhard.87}.
In this method, the GOA is used to reduce the Hill-Wheeler equation
of the GCM to a Schr\"odinger equation for the collective wave function,
the so-called Collective Schr\"odinger Equation (CSE) \begin{equation}
\hat{{\cal \mathcal{H}}}_{coll}\phi_{\alpha}(Q_{3})=\epsilon_{\alpha}\phi_{\alpha}(Q_{3}),\label{SCHCOL}\end{equation}
 where the collective Hamiltonian $\hat{{\cal H}}_{coll}$ is given
by 
\begin{eqnarray}
\hat{{\cal \mathcal{H}}}_{coll} & = & -\frac{1}{\sqrt{G(Q_{3})}}
\frac{\partial}{\partial Q_{3}}\sqrt{G(Q_{3})}\frac{1}{2B(Q_{3})}\frac{\partial}{\partial Q_{3}}
\label{HCOLL}\\
 & + & V(Q_{3})-\epsilon_{0}(Q_{3}).
\end{eqnarray}
 In this expression $G(Q_{3})$ is the metric, $B(Q_{3})$ is the
mass parameter associated with the collective motion along $Q_{3}$,
$V(Q_{3})$ is the collective potential given by the HFB energy 
$V(Q_{3})=\langle\psi(Q_{3})|\hat{H}|\psi(Q_{3})\rangle$
and $\epsilon_{0}(Q_{3})$ is the Zero Point Energy (ZPE) correction.
The eigenfunctions $\phi_{\alpha}(Q_{3})$ of Eq. (\ref{SCHCOL})
have to be normalized to one with the metric $G(Q_{3})$ \begin{equation}
\int dQ_{3}\,\sqrt{G(Q_{3})}\,\phi_{\alpha}^{*}(Q_{3})\phi_{\beta}(Q_{3})=\delta_{\alpha,\beta}\end{equation}
to preserve the hermiticity of $\hat{{\cal \mathcal{H}}}_{coll}$.

It should be mentioned that a CSE can
also be obtained from the Adiabatic Time Dependent Hartree- Fock (ATDHF)
theory \cite{Baranger.78,Brink.76,Villars.77} after quantization
of the semi-classical Hamiltonian for the slow moving collective degrees
of freedom. The collective Hamiltonian obtained in this way has the
same functional form as the GCM+GOA one, but the expression of the
collective parameters is different. Later we will discuss how to
choose these collective parameters.

An interesting characteristic of the collective Hamiltonian for the
octupole degree of freedom is that $\hat{{\cal H}}_{coll}$ is invariant
under the exchange $Q_{3}\rightarrow-Q_{3}$ and, therefore, it is
possible to classify its eigenfunctions, $\phi_{\alpha}(Q_{3})$,
according to their parity under the $Q_{3}\rightarrow-Q_{3}$ exchange.
It is easy to see that the parity of the collective wave function
under the $Q_{3}\rightarrow-Q_{3}$ exchange corresponds to the spatial
parity operation in the correlated wave function built up from $\phi_{\alpha}$.
The inclusion of octupole correlations immediately restores the parity
symmetry lost at the mean field level. Therefore, the solution of
the CSE Eq. (\ref{SCHCOL}) allows
the calculation of the $0^{+}-1^{-}(3^{-})$ energy splitting and
the B(E1) and B(E3) transition probabilities connecting them. At this
point it has to be pointed out that in the present framework where
only time reversal invariant wave functions are considered it is only
possible to describe excited states with average angular momentum
zero. To deal with genuine $1^{-}$ or $3^{-}$states cranking model
wave functions should be considered, which is out of the scope of
the present work. Here we will assume that the cranking rotational
energy of the $1^{-}$state is much smaller than the excitation energy
of the negative parity band head and therefore can be safely neglected.
Also the impact of the cranking term in the transition probabilities
to be discussed next is neglected. With these approximation in mind,
the reduced transition probabilities from the lowest $1^{-}$ and
$3^{-}$ states to the $0^{+}$ ground state can be computed within
the Rotational Model approximation as 
\begin{equation}
B(E\lambda,I_{f}\rightarrow I_{i})=e^{2}\langle I_{i}K\lambda0|I_{f}K\rangle^{2}|
\langle\varphi_{i}|r^{\lambda}Y_{\lambda,0}|\varphi_{f}\rangle|^{2},
\end{equation}
 where $|\varphi_{i}\rangle$ and $|\varphi_{f}\rangle$ are correlated
wave functions obtained in the spirit of the GCM from the collective
wave functions $\phi_{\alpha}(Q_{3})$. The preceding formula can be reduced
to an expression involving those collective wave functions $\phi_{\alpha}(Q_{3})$
by means of the GOA \cite{Nerlo.87}. The final result for $K=0$
bands reads 
\begin{equation}
B(E1,1^{-}\rightarrow0^{+})=\frac{e^{2}}{4\pi}
\left|\langle\phi_{0^{-}}|D_{0}|\phi_{0^{+}}\rangle_{\textrm{COLL}}\right|^{2}
\label{BE1}
\end{equation}
for the E1 electric transition and 
\begin{equation}
B(E3,3^{-}\rightarrow0^{+})=\frac{e^{2}}{4\pi}
\left|\langle\phi_{0^{-}}|Q_{30}(\textrm{PROT)}|\phi_{0^{+}}\rangle_{\textrm{COLL}}\right|^{2},
\label{BE3}
\end{equation}
for the E3 one. In the preceding formulas we have introduced the collective
matrix element of an operator $\hat{O}$ as 
$$
\langle\phi_{0^{-}}|\hat{O}|\phi_{0^{+}}\rangle_{\textrm{COLL}}=
\int dQ_{3}\, G^{1/2}\,\phi_{0^{-}}^{*}(Q_{3})O(Q_{3})\phi_{0^{+}}(Q_{3})
$$
where $O(Q_{3})=\langle\psi(Q_{3})|\hat{O}|\psi(Q_{3})\rangle$.
In Eq. (\ref{BE1}) $D_{0}$ is the dipole moment operator
whose mean value is defined as the difference between the center of
mass of protons and neutrons 
\begin{equation}
D_{0}(Q_{3})=\frac{N}{A}\langle\psi(Q_{3})|\hat{z}_{prot}|\psi(Q_{3})\rangle-
\frac{Z}{A}\langle\psi(Q_{3})|\hat{z}_{neut}|\psi(Q_{3})\rangle.
\label{eq:dipole}
\end{equation}
Finally, $Q_{30}(\textrm{PROT)}$ is the part of the octupole operator
acting on proton's space.

To carry out the collective calculations it is necessary to specify
the collective parameters $G(Q_{3})$, $B(Q_{3})$ and $\epsilon_{0}(Q_{3})$
appearing in the definition of $\hat{\mathcal{H}}_{coll}$ Eq. (\ref{HCOLL}).
As it was said before, there are two sets of parameters coming
from the GCM+GOA and the ATDHF derivation of the collective Hamiltonian.
The set of parameters used in this calculation is an admixture of
the two and it is known as the ATDHF+ZPE set. It includes the mass
parameter $B(Q_{3})$ coming out from the semi-classical Hamiltonian
of the ATDHF theory, the metric of the GCM+GOA and the ZPE correction
computed with the GCM+GOA formula but using the ATDHF mass instead,
i.e. 
\begin{equation}
\epsilon_{0}(Q_{3})=\frac{1}{2}G(Q_{3})B(Q_{3})_{ATDHF}^{-1}.\label{ZPE}
\end{equation}
 This set of parameters was devised to put together the advantages
of the ATDHF set (time-odd components included in the mass term) and
the ones of the GCM+GOA (ZPE correction). This method can be somewhat
justified in the context of the extended GCM
\cite{Reinhard.87,Villars.75} and has been extensively used \cite{Berger.84,Egido.89}.

The calculation of the collective parameters involves the inversion
of the HFB stability matrix which is closely related to the matrix
of the RPA equation. At present, this is a formidable task and approximations
are needed. The approximation used in this article -- called {}``cranking
approximation\textquotedbl{} \cite{Reinhard.78,Girod.79} -- neglects
the off-diagonal terms of the stability matrix allowing to invert
it analytically but at the cost of including the two body interaction
only through the mean field. Although this approximation has been
extensively used in the literature for the calculation of collective
masses and moments of inertia (see, for instance, Refs. \cite{Baran.81,Berger.84,Boning.85})
its validity has not been properly established. Using the {}``cranking
approximation'', the ATDHF+ZPE parameters are given by 
\begin{equation}
G(Q_{3})=\frac{M_{-2}(Q_{3})}{2M_{-1}^{2}(Q_{3})},\,\,\, B(Q_{3})=\frac{M_{-3}(Q_{3})}{M_{-1}^{2}(Q_{3})};
\label{eq:COLLM}
\end{equation}
where the quantities $M_{-n}(Q_{3})$ ($n=1,2,3$ ) are defined as
\begin{equation}
M_{-n}(Q_{3})=\sum_{k,l}\frac{\left|\left(Q_{30}\right)_{kl}^{20}\right|^{2}}{(E_{k}+E_{l})^{n}}.
\label{eq:Mn}
\end{equation}
In the preceding expression, $E_{k}$ are the quasi-particle energies
and $\left(Q_{30}\right)_{kl}^{20}$ are the matrix elements of the
$20$ part \cite{Ring.80} of the octupole operator $\hat{Q}_{30}$
in the quasi-particle basis of the HFB wave function $|\psi(Q_{3})\rangle$.
This form of the collective mass is usually referred to in the literature
as Belyaev-Inglis mass \cite{Ring.80}.

\section{Results}

In the subsequent subsections the results obtained with the BCP1 \cite{Baldo.08}
functional and regarding octupole properties of some radium and barium
isotopes will be discussed. The other functional defined in \cite{Baldo.08}
and referred to as BCP2 will not be explicitly considered here although
the calculations were carried out for that case too. The reason is
the strong similarities between BCP1 and BCP2 results that produced most
of the curves one on top of another, making it impossible to
differenciate them in the plots presented.

\subsection{Low excitation energy properties in the radium isotopes}

\begin{figure}
\includegraphics[width=0.95\columnwidth]{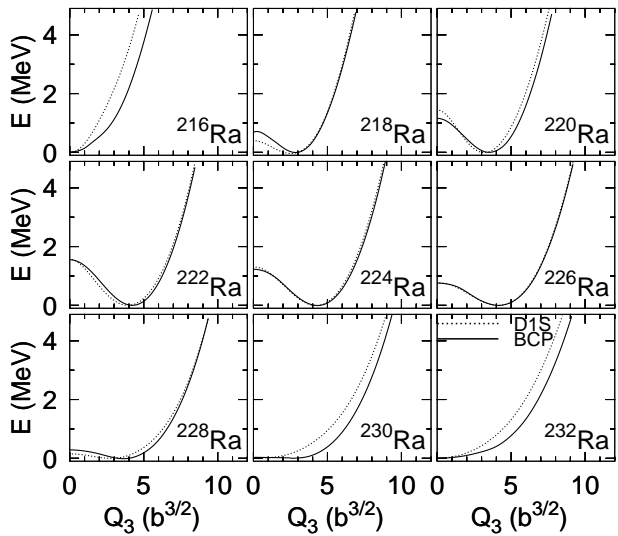}%
\caption{The HFB mean field energy as a function of the octupole moment $Q_{3}$
(in units of $b^{3/2}=10^{3}\textrm{ fm}^{3}$) for the isotopes of
radium (Z=88) from A=216 up to A=232. Results for both the BCP1 energy
density functional (full curves) and the Gogny D1S force (dotted curves)
are shown.\label{fig:MFPES_Ra}}

\end{figure}

Octupole deformation properties of the radium isotopes were the first
to be addressed from a microscopic point of view with the Gogny force,
first at the mean field level \cite{Egido.89} and next including
exact restoration of the parity symmetry \cite{Egido.91}. At the
mean field level, the first quantity to analyze is the potential energy
curve (PEC) as a function of the octupole moment that determines both
the ground state minimum and its stiffness. Let us point out that for
every point in the PEC the other multipole moments (quadrupole, hexadecapole,
etc) are selfconsistently determined as to produce the lowest energy.
The PECs computed with the BCP1 \cite{Baldo.08} energy density
functional and the Gogny D1S \cite{Decharge.80,Berger.84} force
are depicted in Fig. \ref{fig:MFPES_Ra}. As can be seen in the plot,
the results for the two types of interactions look very similar in
all the nuclei considered. It is observed how, whenever a minimum
appears (in the nuclei from $^{218}$Ra to $^{228}$Ra) in the Gogny
D1S calculation at an octupole deformation different from zero, the same happens
and at the same $Q_{3}$ value in the BCP1 calculation. For the nuclei
with the minimum at $Q_{3}=0$ the Gogny D1S force shows a tendency
to produce a stiffer parabolic behavior in the PEC than in BCP1. The
depth of the octupole deformed minima is also very similar for both
kinds of calculations and reaches its maximum value of 1.5 MeV for
the nucleus $^{222}$Ra which therefore can be considered as the strongest
octupole deformed nuclei of the considered chain.

\begin{figure}
\includegraphics[width=0.95\columnwidth]{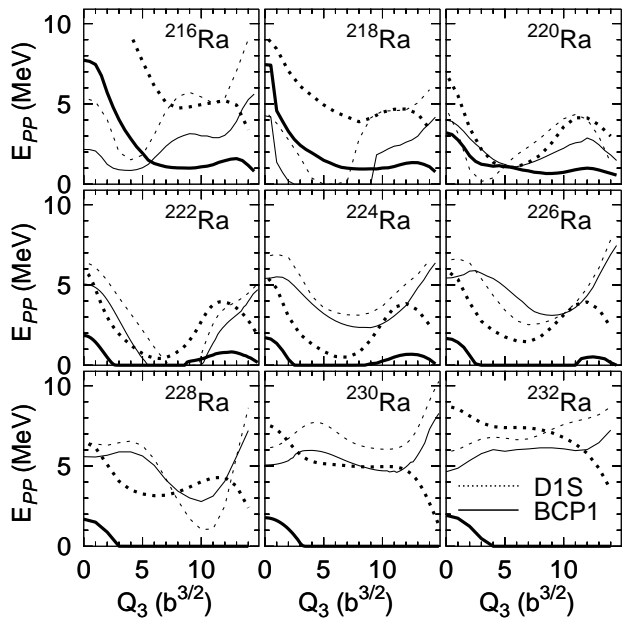}%
\caption{The particle-particle (pp) correlation energy defined as $E_{pp}=\frac{1}{2}\textrm{Tr}\Delta\kappa$
is plotted as a function of the octupole moment $Q_{3}$ (in units
of $b^{3/2}=10^{3}\textrm{ fm}^{3}$) for the radium isotopes considered.
Dotted lines correspond to the pp correlation energies whereas full
lines are meant to represent the same quantity for protons. Thick
lines are used to depict the results of the calculations with the
BCP1 functional whereas thin lines are for the results with the Gogny
D1S force.\label{fig:PAIRING_Ra}}

\end{figure}

In Fig \ref{fig:PAIRING_Ra} we show the particle-particle correlation
energy defined as $E_{pp}=\frac{1}{2}\textrm{Tr}\Delta\kappa$ and
given in terms of the usual pairing field $\Delta$ and pairing tensor
$\kappa$ of the HFB method. This quantity gives a rough idea of the
amount of pairing correlations in the system. It can also be used
as an indicator of the size of the single particle level density around
the Fermi surface as strong pairing correlations are direct consequence
of high level densities. This energy is also correlated with the pairing
gap that represents the energy of the lowest two quasi-particle excitation
and therefore it is closely related to the collective inertias to
be discussed below. The overall tendency of $E_{pp}$ is to decrease
with increasing octupole moment up to values of $Q_{3}=10$ b$^{3/2}$which
correspond to typical excitation energies of 5-6 MeV above the ground
state in the PECs. From there on we observe, depending on the nucleus,
stationary behaviors or mild increases. We also notice that the $E_{pp}$
computed with BCP1 are greater than the ones computed with D1S for
the light isotopes $^{216}$Ra and $^{218}$Ra and for the two species
of nucleons. For the nucleus $^{220}$Ra the particle-particle correlation
energies for protons and neutrons are similar in both calculations
and from there on and up to the $^{230}$Ra isotope the D1S correlation
energies are larger than the BCP1 ones. For the heaviest isotope considered
$^{232}$Ra the $E_{pp}$ energy for neutrons is larger for BCP1 than
for D1S and the opposite is true for protons. 

\begin{figure*}
\includegraphics[width=1.5\columnwidth]{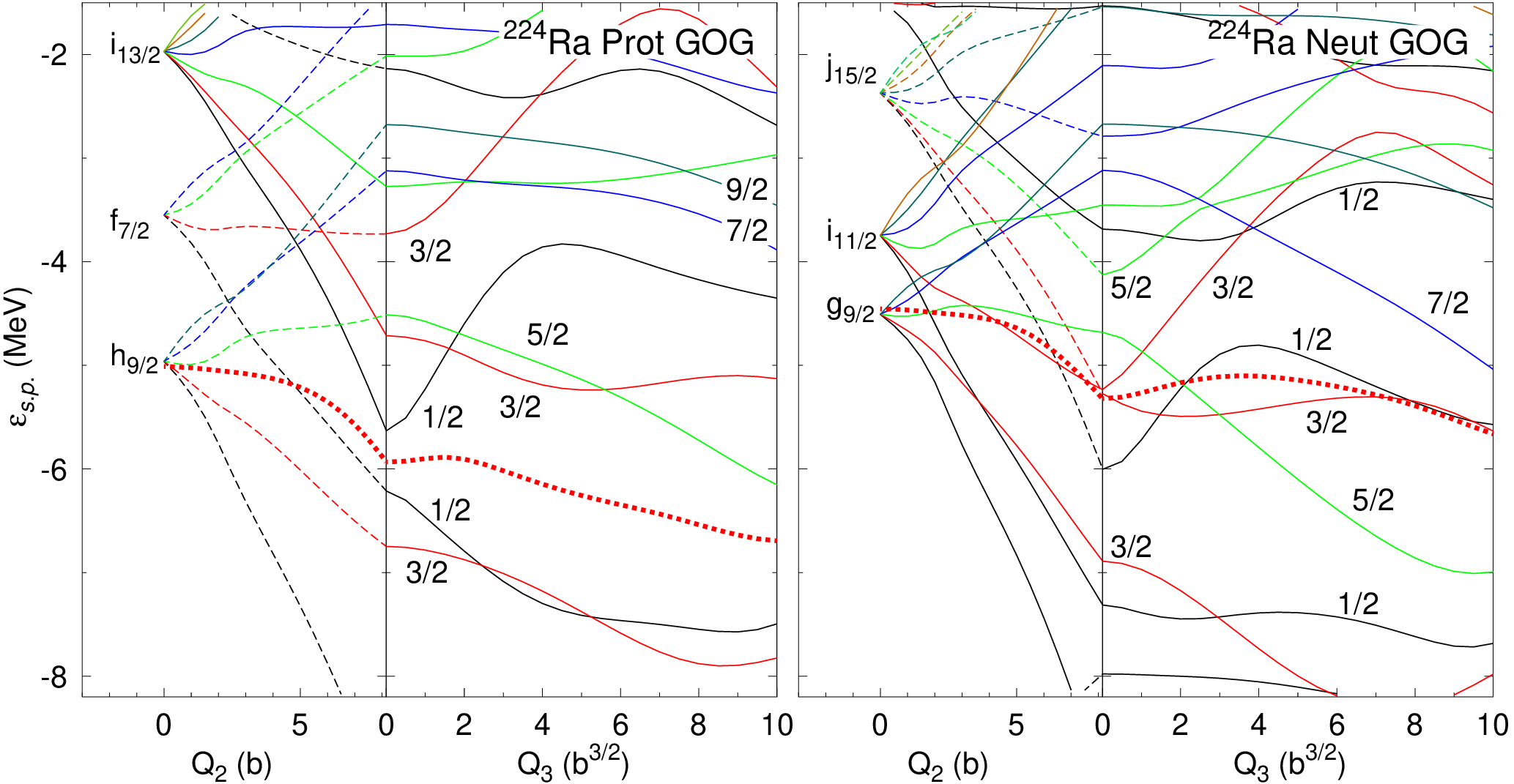}

\includegraphics[width=1.5\columnwidth]{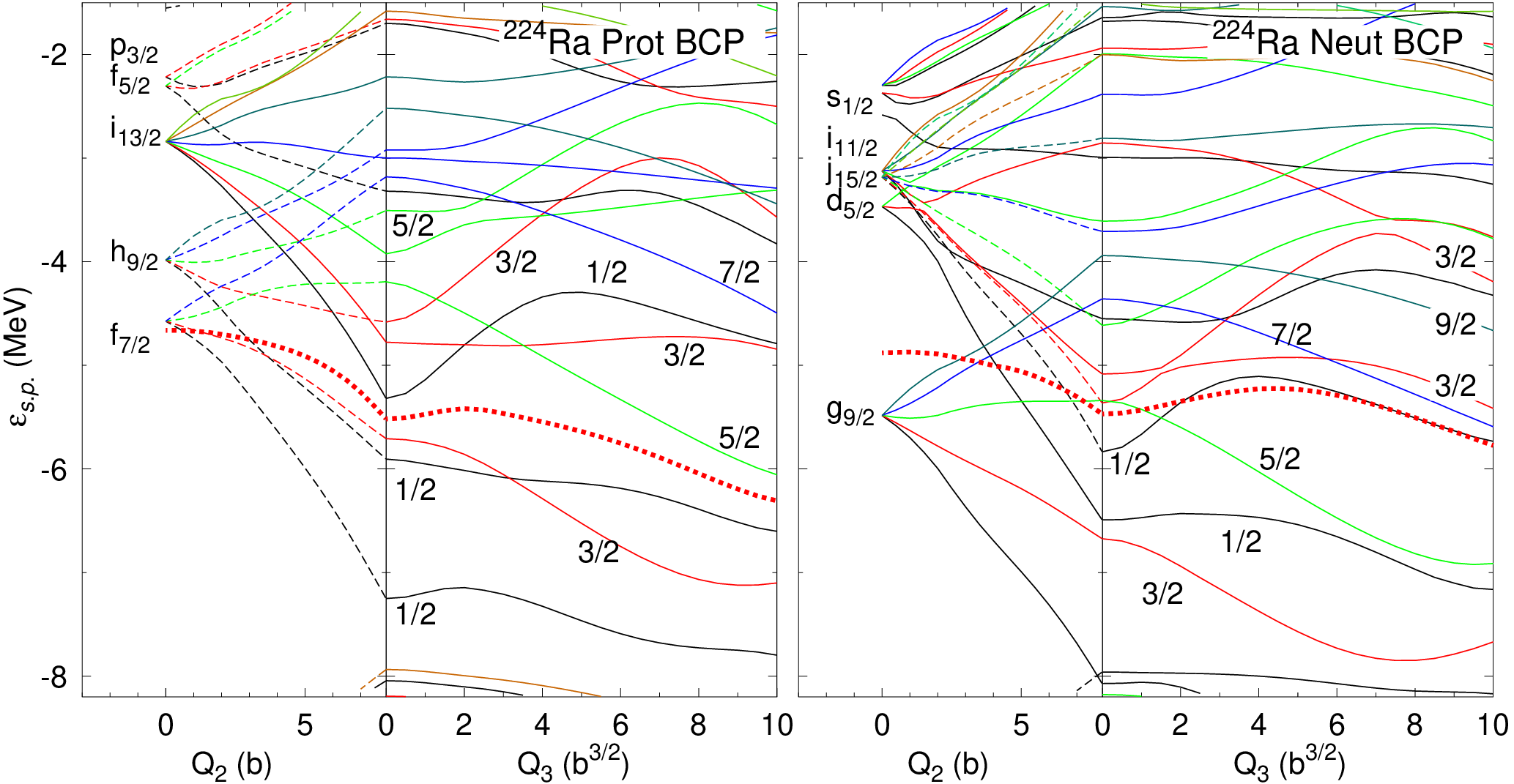}

\caption{(Color online) Single particle energies (SPE) of the nucleus $^{224}$Ra
both for protons (left panels) and neutrons (right panels) and the
two interactions/functionals used in the calculations (Gogny D1S in
the upper panels, BCP1 in the lower ones). In each panel, the single
particle energies are plotted first as a function of the mass quadrupole
moment $Q_{2}$ starting at sphericity ($Q_{2}=0$ b) up to the value
corresponding to the self-consistent minimum in the quadrupole variable
at $Q_{3}=0$ b$^{3/2}$ ($Q_{2}$=8.12 b for the Gogny calculation
and 7.88 b for the BCP1 one). In this plot the spherical quantum numbers
obtained at $Q_{2}=0$ b are given. On the right hand side of each
panel, the SPE as a function of the octupole moment starting at $Q_{3}=0$
b$^{3/2}$are plotted. The octupole deformed self-consistent minimum
is locate in this nucleus at around $Q_{3}=4$ b$^{3/2}$. Thick dotted
lines indicate the Fermi level. As a consequence of axial symmetry,
the $K$ quantum number of each level remains fixed along the plot. Each
$K$ value has an associated color (black, $K=1/2$; red, $K=3/2$ ;
green, $K=5/2$; blue, $K=7/2$; dark green, $K=9/2$, etc). The parity
of each state when $Q_{3}=0$ (i.e., in the $Q_2$ plot) is determined 
by the kind of line: full for positive parity and dashed for negative parity.
Finally, the $K$ quantum numbers of specific levels around the Fermi
level are given.
\label{fig:SPE224Ra}}

\end{figure*}

In Fig. \ref{fig:SPE224Ra} the single particle energies (SPE) are
plotted as a function of the octupole moment $Q_{3}$ for both protons
and neutrons. Results of both calculations are shown in different
panels ( Gogny D1S upper and BCP1 lower panels). In all the cases,
the Fermi level is represented by a thick dotted line. Once the stretching
effect of the bigger effective mass of BCP1 (1, versus 0.7 in Gogny
D1S) is accounted for, the similitude between the
single particle energies obtained with Gogny D1S and BCP1 around the
Fermi level and $Q_{3}$ values in the neighborhood of the mean field
minimum is remarkable. It is also possible to say that the behavior of most of the
levels as a function of both $Q_{2}$ and $Q_{3}$ is quite similar
in both kinds of calculations. This is so in spite of the different
ordering of the spherical orbitals: for protons the $f_{7/2}$ and
$h_{9/2}$ are reversed in the spectrum of Gogny D1S as compared to
the spectrum of BCP1. For neutrons there is also such an inversion between
the $j_{15/2}$ orbital and the $i_{11/2}$ orbital and also the separation
between the $i_{11/2}$ and $g_{9/2}$ orbitals is much larger in
BCP1 than in Gogny D1S. That the single particle spectrum looks similar
in the region of well developed quadrupole deformation and also as
a function of octupole deformation is probably a consequence of the
collective character of those collective degrees of freedom where
the geometry of the shape of the nucleus is more important than quantum
mechanical effects. To make the argument more quantitative, we have 
analyzed the structure of the single particle wave functions in terms
of Nilsson quantum numbers and found that the levels around the Fermi
surface have similar structures. A typical example for protons is the
$K=5/2^-$ level that for $Q_3=0$ lies at -4.5 MeV  in the D1S case and at
-4.2 MeV  in the BCP one. This level originates in the $h_{9/2}$ spherical
orbital in the D1S case whereas it comes from a $f_{7/2}$ in the BCP one.
The Nilsson quantum numbers for the D1S orbital are $[523]$ (66$\%$), 
$[532]$ (11$\%$), $[503]$ (7$\%$), $[743]$ (6$\%$), and other small components
 whereas for BCP1 they are  $[523]$ (46$\%$),  $[532]$ (15$\%$), $[503]$ 
 (9$\%$), $[312]$ (7$\%$) and smaller components. We can also consider another
 example in the neutron side where, at $Q_3=0$ there is a $K=3/2^+$ orbital
 at around -5 MeV that originates from a $i_{13/2}$ spherical level
 in the D1S calculations and from a $d_{5/2}$ in the BCP one. The
 Nilsson quantum numbers obtained are $[631]$ (31$\%$), $[642]$ (25$\%$),
 $[611]$ (15$\%$), $[862]$ (8$\%$) and small components for D1S and
 $[642]$ (47$\%$), $[631]$ (16$\%$), $[862]$ (13$\%$), $[422]$ (7$\%$) 
 and small components for BCP. From the preceding examples and other orbitals
 considered (but not displayed here) we conclude that the quantitative
 structure of the levels is quite similar in the two calculations 
 irrespective of their spherical origin. This reinforces our suggestion
 about the fundamental role played by the collective degrees of freedom
 in the determination of single particle wave functions.

The conditions for the development of octupolarity
are clearly satisfied in this nucleus as can be seen in the single
particle plot: for protons there are {}``$f_{7/2}$'' levels below
the Fermi level with $K=1/2$ and 3/2 and at the same time $i_{13/2}$
orbitals with $K=1/2$ and $3/2$ are just above the Fermi level.
The same happens in the neutron case with the $g_{9/2}$ orbital well
below the Fermi level and the $j_{15/2}$ with $K=1/2$ and 3/2 at
the Fermi level (please remember the super-fluid character of neutrons
that makes the Fermi level concept a diffuse one). Another condition
for the development of a minimum is the presence of a region of low
level density in the SPE spectrum (Jahn-Teller effect, see Ref. \cite{Bohr.75}
for a general discussion in the nuclear context). We observe in the
two proton spectra in Fig. \ref{fig:SPE224Ra} how the Fermi level
of protons lies in the middle of a low level density region at $Q_{3}=4$
b$^{3/2}$ which corresponds to the position of the minimum. For neutrons
and around $Q_{3}=4$b$^{3/2}$ we also observe a region of low level
density near the Fermi level which is more pronounced for the BCP1
results. As the number of neutrons is increased the Fermi level moves
upward and enters a region of high level density that is unable to
lead to a deformed minimum as is the case for $^{230}$Ra and heavier
isotopes.
 
Finally, we would like to mention that the differences observed
in the position of the single particle levels in Fig. \ref{fig:SPE224Ra}
has little impact on the quantum numbers of neighboring odd-A nuclei as
in the present mean field framework those quantum numbers have to be
obtained after a selfconsistent blocking mean field procedure and it is
not enough to block the single particle orbitals of Fig. \ref{fig:SPE224Ra}
as would be the case with a description based on  a Nilsson diagram. 
Work to implement such blocking mechanism in the BCP case is under way
and will be reported in the near future.

\begin{figure}
\includegraphics[width=0.95\columnwidth]{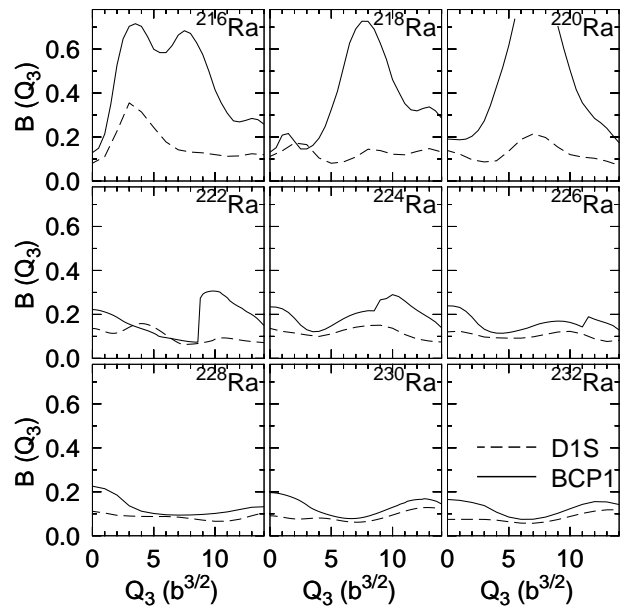}%
\caption{The octupole collective inertia parameter $B(Q_{3})$ entering the
one dimensional collective Schr\"odinger Hamiltonian (see text for details)
is shown as a function of the octupole moment $Q_{3}$ (in units of
$b^{3/2}=10^{3}\textrm{ fm}^{3}$) for the radium isotopes considered. 
Full lines are for the calculations with the BCP1 functional and
dashed ones for the results obtained with the Gogny D1S force.\label{fig:CollM_Ra}}

\end{figure}

In Fig. \ref{fig:CollM_Ra} we show the collective inertia $B(Q_{3})$
associated with the octupole degree of freedom (see Eqs. (\ref{eq:COLLM})
and (\ref{eq:Mn})) and playing a central
role in the collective Schr\"odinger Hamiltonian of the previous section. 
As a consequence of the presence in its definition
of a denominator with powers of the two quasiparticle energies $E_{\mu}+E_{\nu}$,
the collective inertia is roughly speaking inversely proportional
to the amount of pairing correlations (the pairing gap to be more
quantitative) and directly proportional to the effective mass of the
interaction. The lower pairing correlations present in BCP1 are not
able to compensate for the higher effective mass and as a consequence
the BCP1 inertias are higher than the Gogny D1S ones. Thus the energies
obtained as a solution of the one dimensional collective Hamiltonian,
which are roughly speaking proportional to the inverse of the square
root of the collective mass (remember the standard harmonic oscillator
formula relating the oscillator's frequency $\omega$ with the spring
constant and the mass $\omega=\sqrt{k/m}$), are expected to reach
lower values for BCP1 than for D1S. It is also worth noticing that
the peaks observed in the $B(Q_{3})$ plots are related to regions
of low pairing correlations as is easily deduced by comparing Fig.
\ref{fig:CollM_Ra} with Fig. \ref{fig:PAIRING_Ra}.

\begin{figure}
\includegraphics[width=0.95\columnwidth]{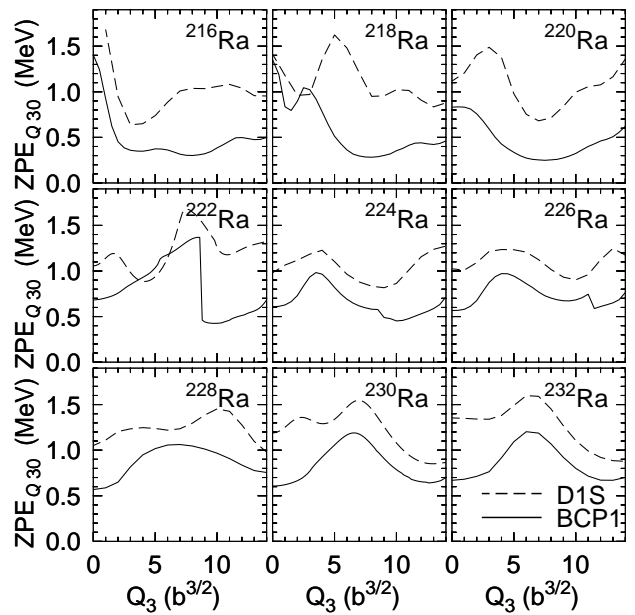}\caption{The octupole zero point energy correction 
$\epsilon(Q_{3})$ of Eq.
\ref{ZPE} is plotted as a function of the octupole moment $Q_{3}$
(in units of $b^{3/2}=10^{3}\textrm{ fm}^{3}$) for the isotopes of
radium considered. Full lines are for the calculations with the BCP1
functional and dashed ones for the results obtained with the Gogny
D1S force.\label{fig:ZPE_Ra}}

\end{figure}

In Fig \ref{fig:ZPE_Ra} the zero point energy correction $\epsilon(Q_{3})$
of Eq. \ref{ZPE} is given for the isotopes of radium considered and
the BCP1 functional and the Gogny D1S force. The values of $\epsilon(Q_{3})$
are correlated with the inverse of the collective inertia $B(Q_{3})$,
as can easily be noticed by comparing Figs \ref{fig:CollM_Ra} and
\ref{fig:ZPE_Ra}. The range of variation is typically of around half
an MeV in the interval of interest between $Q_{3}=0$b$^{3/2}$ and
$Q_{3}\approx5$b$^{3/2}$ and most of the nuclei considered although
there are exceptions like the nucleus $^{216}$Ra. The effect of the
ZPE is to increase the depth of the octupole well for the lighter
nuclei $^{216-220}$Ra whereas it is the opposite in all the heavier
isotopes. The impact of this effect on the properties of the solutions
of the CSE is not as pronounced as it
could be thought because of the effect of the collective masses (correlated
to the behavior of the ZPE) that tends to cancel out the one of the
ZPE.

\begin{figure}
\includegraphics[width=0.95\columnwidth]{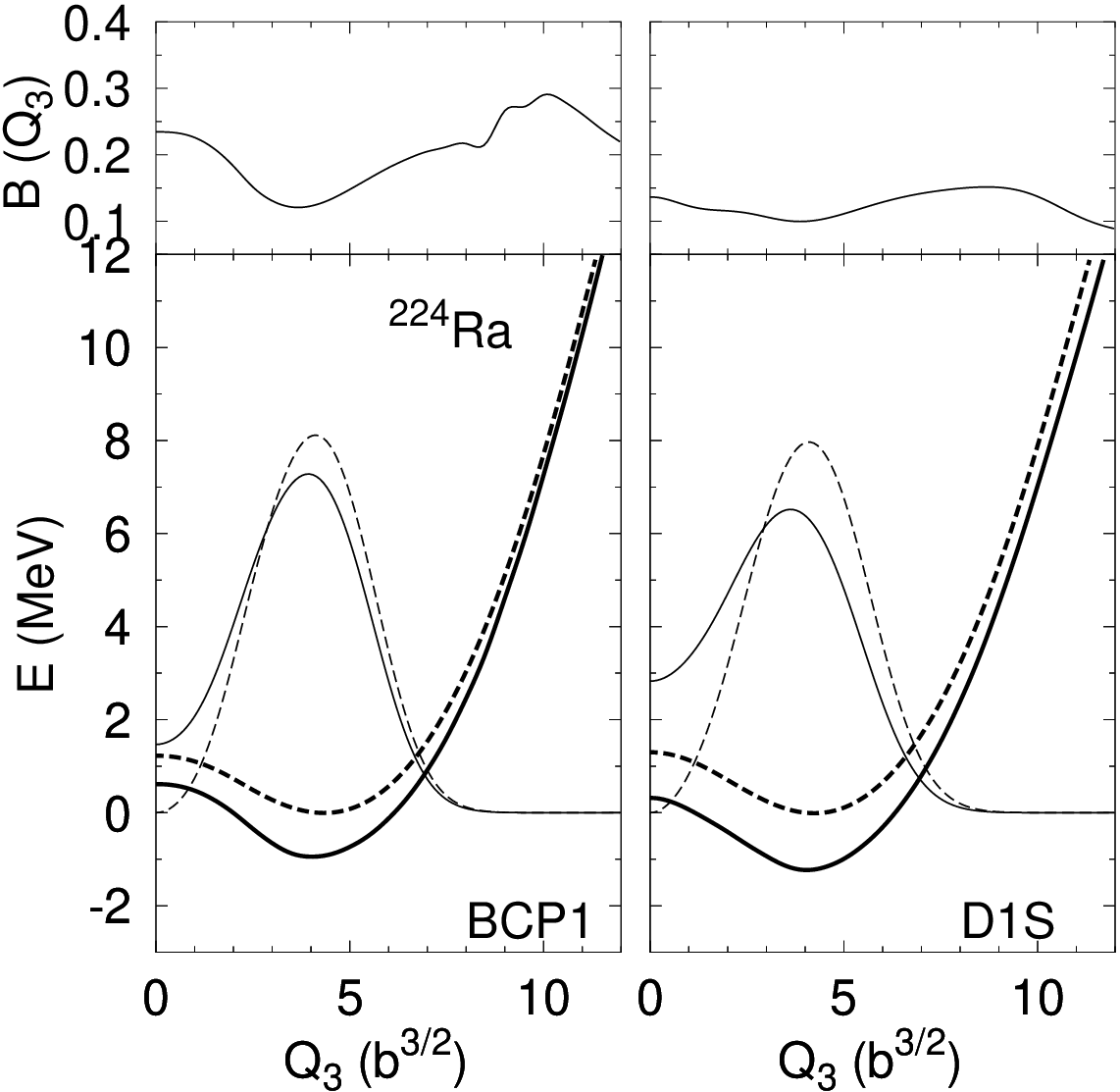}

\caption{The relevant quantities entering the Collective Schr\"odinger equation
for the two cases considered, namely the calculation with the Gogny
D1S force (right hand side panels) and the calculation with the BCP1
energy density functional (left hand side panels). In the upper panels,
the octupole collective mass $B(Q_{3})$ (already shown in Fig. \ref{fig:CollM_Ra})
is depicted as a function of the octupole moment $Q_{3}$ (in units
of $b^{3/2}=10^{3}\textrm{ fm}^{3}$). In the lower panels, the thick
lines represent the HFB energy (dashed line) and the CSE potential
energy (HFB energy minus zero point energy, full line). The thin lines
correspond to  the square of the collective wave functions corresponding
to the lowest positive parity state $|\phi_{0^{+}}(Q_{3})|^{2}$ (full
line) and lowest energy negative parity state $|\phi_{0^{-}}(Q_{3})|^{2}$
(dashed line).\label{fig:CSE}}

\end{figure}

\begin{figure*}
\includegraphics[width=1.75\columnwidth]{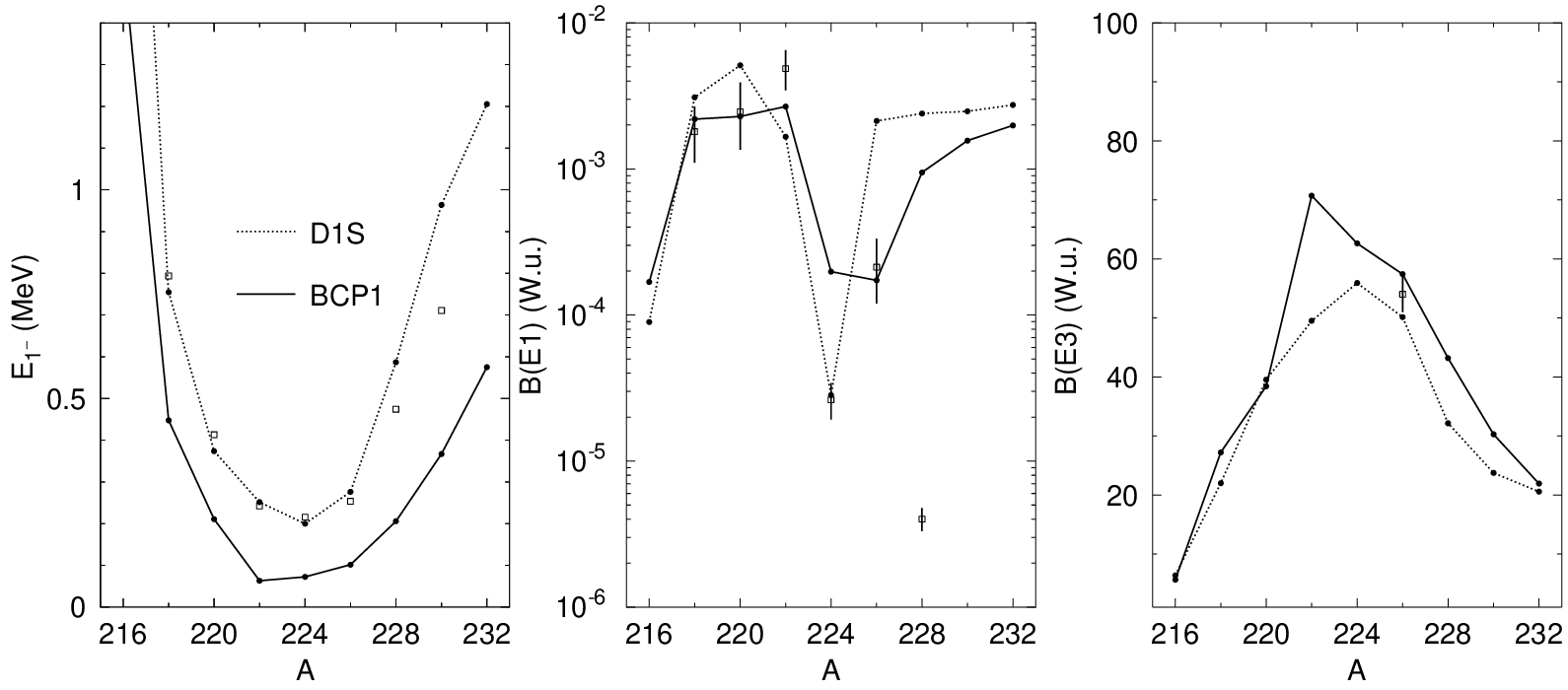}

\caption{In the left hand side panel, the energies of the $1^{-}$states, obtained
by solving the 1D collective Schr\"odinger Hamiltonian, are displayed
as a function of the mass number for the radium isotopes considered.
Theoretical results (lines, full for the results obtained with the
BCP1 functional and dotted for the results obtained with the D1S force)
are plotted along with the experimental data (boxes). In the middle
panel the $B(E1,1^{-}\rightarrow0^{+})$ transition probabilities
in W.u. are given as a function of the mass number for the radium
isotopes considered. Finally, in the right hand side panel the $B(E3,3^{-}\rightarrow0)$
in W.u. is given for the different radium isotopes considered. \label{fig:RES_Ra}}

\end{figure*}

\begin{figure}
\includegraphics[width=0.95\columnwidth]{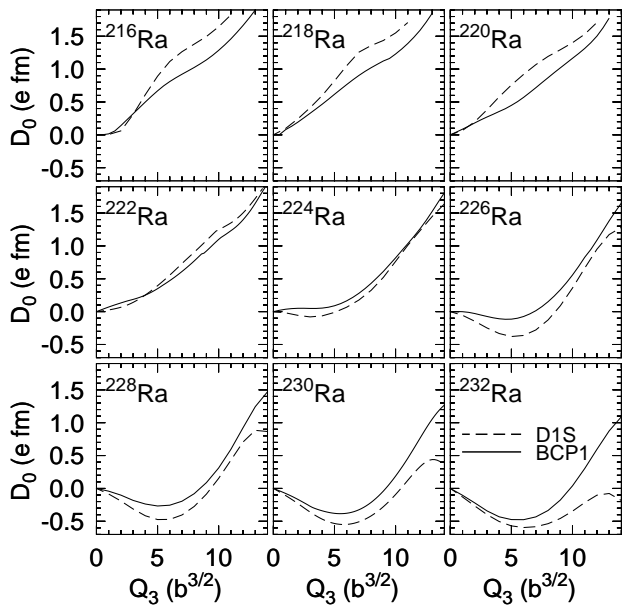}

\caption{Dipole moments $D_{0}$(fm) are represented as a function of the octupole
moment $Q_{3}$ (in units of $b^{3/2}=10^{3}\textrm{ fm}^{3}$) for
the nine isotopes of Ra considered. Dashed lines correspond to the
results obtained with the Gogny D1S force whereas the full line ones
are for the results obtained with the BCP1 functional. \label{fig:Dipole-moments-Ra}}

\end{figure}

With the potential energy curve, the collective mass and the zero
point energy correction, all the ingredients needed to solve the CSE are at 
our disposal. In Fig. \ref{fig:CSE}
we have shown all those ingredients together in two plots corresponding
to the results with the Gogny D1S force and BCP1 functional. In each
of the plots we have depicted in the lower panel the HFB energy curve
(dashed line) as a function of the octupole moment $Q_{3}$ and shifted
to put the minimum at zero energy. The full curve closely following
the dotted one is the potential energy entering the CSE that is obtained
by subtracting the ZPE energy correction to the HFB energy. As observed
in the plot the collective potential energy is rather similar to the
HFB energy. Also the HFB energies obtained with the D1S force and
the BCP1 functional  calculations are rather similar. In the same
panel, the square of the collective amplitudes $|\phi_{\alpha}(Q_{3})|^{2}$
for the lowest lying state of each parity are plotted. The negative
parity amplitudes look rather similar in both calculations but this
is not the case for the positive parity amplitude which is higher
around $Q_{3}=0$ for Gogny D1S than for BCP1. The different behavior
of the positive parity amplitude is related to the different collective
masses obtained in both calculations and given in the upper panels.
The Gogny D1S collective mass is much lower around $Q_{3}=0$ than
the one obtained with BCP1 and, as discussed in Ref. \cite{Egido.89,Egido.90},
this enhances the collective amplitude around that value. As a consequence
of the lower mass obtained with the D1S force the energy of the $1^{-}$state
computed after solving the CSE is higher (200 keV) than the one obtained
with the BCP1 functional (73 keV). On the other hand, the effect on
the $B(E1,1^{-}\rightarrow0^{+})$ and $B(E3,3^{-}\rightarrow0^{+})$
transition probabilities is to yield smaller values for D1S than for
BCP1 as the overlap between the positive and negative parity amplitudes
is smaller in the later case. However, recalling the expression of
the transition probabilities of Eqs. (\ref{BE1}) and (\ref{BE3}),
it is easy to realize the reduced impact of the region around $Q_{3}=0$b$^{3/2}$
on the final quantities as each of the factors of the integrands,
$D_{0}(Q_{3})$ and $\left(Q_{3}\right)_{prot}(Q_{3})$, are zero
for $Q_{3}=0$ b$^{3/2}$.

The energies of the $1^{-}$ states and transition probabilities obtained
by solving the CSE with the collective parameters deduced from the
Gogny D1S and BCP1 calculations are depicted in Fig. \ref{fig:RES_Ra}
along with available experimental values. For the energy of the $1^{-}$
states we observe that both the BCP1 functional and the Gogny D1S
interaction reproduce quite nicely the experimental isotopic trend
(see \cite{Butler.91,Butler.96} and references therein) with a minimum
around A=224. The very good reproduction of the experimental data
in the calculation with Gogny D1S can be considered as accidental
in the sense that the absolute values of the $1^{-}$excitation energies
crucially depend on the amount of pairing correlations (through the
collective mass) which are not so well characterized at the mean field
level. A more robust indicator of the quality of the results is the
reproduction of the isotopic trend which is very good in both, D1S
and BCP1, calculations. Concerning the $B(E1,1^{-}\rightarrow0^{+})$
transition probabilities we observe a pronounced minimum around $A=224$
in the two calculations that is a direct consequence of the behavior
of the dipole moment as a function of the octupole moment for different
isotopes. This dip in the $B(E1,1^{-}\rightarrow0^{+})$ values is
also observed experimentally (see \cite{Butler.91,Butler.96} and
references therein) and is well reproduced by the Gogny D1S force
and reasonably well by the BCP1 functional. On the other hand, BCP1
nicely reproduce the $B(E1)$ of $^{226}$Ra, whereas the Gogny D1S
force result yields a too high value. Concerning the $B(E3,3^{-}\rightarrow0^{+})$
we observe a maximum around $A=224$ which is correlated to the minimum
in the energies of the $1^{-}$ states. Both calculations reproduce
quite well the only experimental value known \cite{Spear.90,Kibedi.02}. 

In order to get a more detailed understanding of the isotopic behavior
of the $B(E1,1^{-}\rightarrow0^{+})$, it is convenient to look at
the behavior of the dipole moment $D_{0}$ as a function of $Q_{3}$
for the different isotopes considered. According to Eq. \ref{BE1},
the value of the $B(E1)$ transition probability is proportional to
the square of the average of the dipole moment over the whole $Q_{3}$interval
and weighted with the product of the ground state positive parity
collective wave function times the lowest negative parity one. The
dipole moment entering Eq. \ref{BE1} is represented as a function
of $Q_{3}$ in Fig. \ref{fig:Dipole-moments-Ra} for the Ra isotopes
studied. Due to the good parity of the $Q_{3}=0$ b$^{3/2}$ solution
the center of mass is located at the origin of coordinates and therefore
the dipole moment is always zero in that case. We observe that at
the beginning of the isotopic chain the dipole moment increases monotonically
as a function of $Q_{3}$but its slope decreases with increasing neutron
number. For $^{224}$Ra and also $^{226}$Ra the slope is almost zero
in the region from $Q_{3}=0$ b$^{3/2}$ and up to $Q_{3}=5$ b$^{3/2}$,
which is the region of interest where the collective wave function
weight is different from zero. As a consequence, it is expected that
the $B(E1)$ has to reach a minimum for one of these isotopes. For
$^{228}$Ra and heavier isotopes the dipole moment in the region of
interest decreases monotonically with a somehow constant slope what
explains the increase in $B(E1)$ in those isotopes as compared to
$^{224}$Ra and $^{226}$Ra as well as their almost constant value
as a function of neutron number. The behavior of $D_{0}$ with neutron
number can be easily understood by looking at Fig. \ref{fig:SPE224Ra}
where the SPE are plotted. There we observe how increasing the number
of neutrons leads to the occupancy of more levels belonging to the
high-j orbitals $i_{11/2}$ and $j_{15/2}$. Due to the high total
angular momentum value j of those orbitals, the spatial distribution
of probability must have regions of large curvature that result in
large values of $\langle z\rangle$ for those orbitals. Thus, increasing
the number of neutrons increases the number of particles in those
orbitals and the value of $\langle z\rangle_{neut}$ also increases,
producing a decrease of $D_{0}$ that is clearly seen in Fig. \ref{fig:Dipole-moments-Ra}.

\subsection{Neutron-rich barium isotopes}

\begin{figure}
\includegraphics[width=0.95\columnwidth]{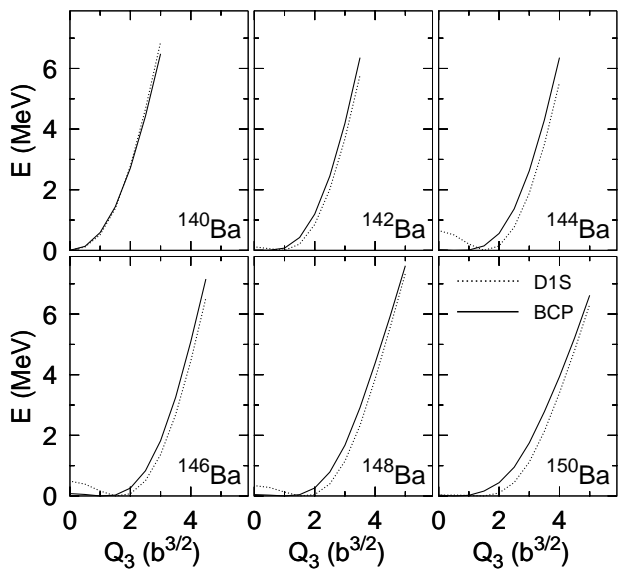}%
\caption{The HFB mean field energies of barium isotopes are plotted as a function
of the octupole moment $Q_{3}$ (in units of $b^{3/2}=10^{3}\textrm{ fm}^{3}$)
for the six even-even isotopes of barium (Z=56) considered.\label{fig:MFPES_Ba}}

\end{figure}

Neutron rich barium isotopes (Z=56) with mass numbers 142, 144 and
146 show several of the characteristics of octupole deformation in
their ground states and yrast bands. Experiments \cite{Phillips.96}
using the fragment yield of the $^{252}$Cf fission decay provided
information on the yrast and negative parity rotational bands in these
nuclei showing the typical alternating parity rotational band pattern
representative of octupole deformed nuclei. For this reason we have
performed calculations for the even-even barium isotopes with atomic
numbers from A=140 up to A=150 to check the predictions of the BCP1
functional concerning octupolarity. Previous calculations with the
Gogny D1S force in this region either at zero spin \cite{Egido.90,Martin.94}
or at high spins using the standard HFB cranking model \cite{Garrote.97}
and even including parity projection \cite{Garrote.98} have been
performed. In all the cases the agreement with experiment was satisfactory.

\begin{figure}
\includegraphics[width=0.95\columnwidth]{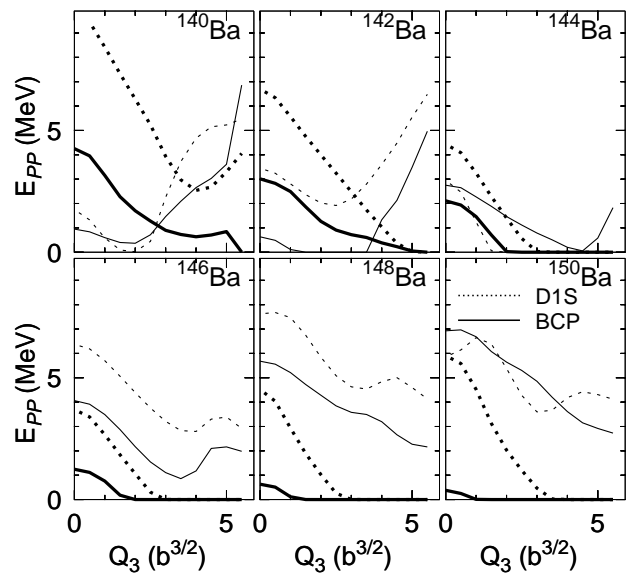}%
\caption{The particle-particle (pp) correlation energy defined as $E_{pp}=\frac{1}{2}\textrm{Tr}\Delta\kappa$
is plotted as a function of the octupole moment $Q_{3}$ (in units
of $b^{3/2}=10^{3}\textrm{ fm}^{3}$) for the barium isotopes considered.
Dotted lines correspond to the pp correlation energies of neutrons
whereas full lines are meant to represent the same quantity for protons.
Thick lines are used to depict the results of the calculations with
the BCP1 functional whereas thin lines are for the results with the
Gogny D1S force.\label{fig:MFPair_Ba}}

\end{figure}

In Fig. \ref{fig:MFPES_Ba} we show for the Gogny D1S force (dotted
line) and BCP1 functional (full line) the PEC corresponding to the
six barium isotopes considered. Whereas in the Gogny D1S predictions
it turns out that three isotopes have an octupole deformed minimum
(namely, $^{144}$Ba, $^{146}$Ba and $^{148}$Ba) this is not the
case for the BCP1 results. However, in those nuclei the PEC calculated
with the BCP1 functional are very flat around the $Q_{3}=0$ b$^{3/2}$
minimum, which is a clear signature of a strong instability in the
octupole degree of freedom. In addition, the depth of the octupole
minima computed with the Gogny D1S force never exceed the 0.7 MeV
found in the case of $^{144}$Ba, which is a quite small height as
compared to the typical energies of the vibrational octupole states.
Therefore, the existence of the octupole minima can not be considered
as conclusive. For the nuclei $^{142}$Ba and $^{150}$Ba the results
of both kinds of calculations show very flat curves around the $Q_{3}=0$
b$^{3/2}$ minimum indicating some degree of instability against the
octupole degree of freedom. Finally, the nucleus $^{140}$Ba is found
to be rather stiff against octupole deformation in the two cases.

\begin{figure*}
\includegraphics[width=1.5\columnwidth]{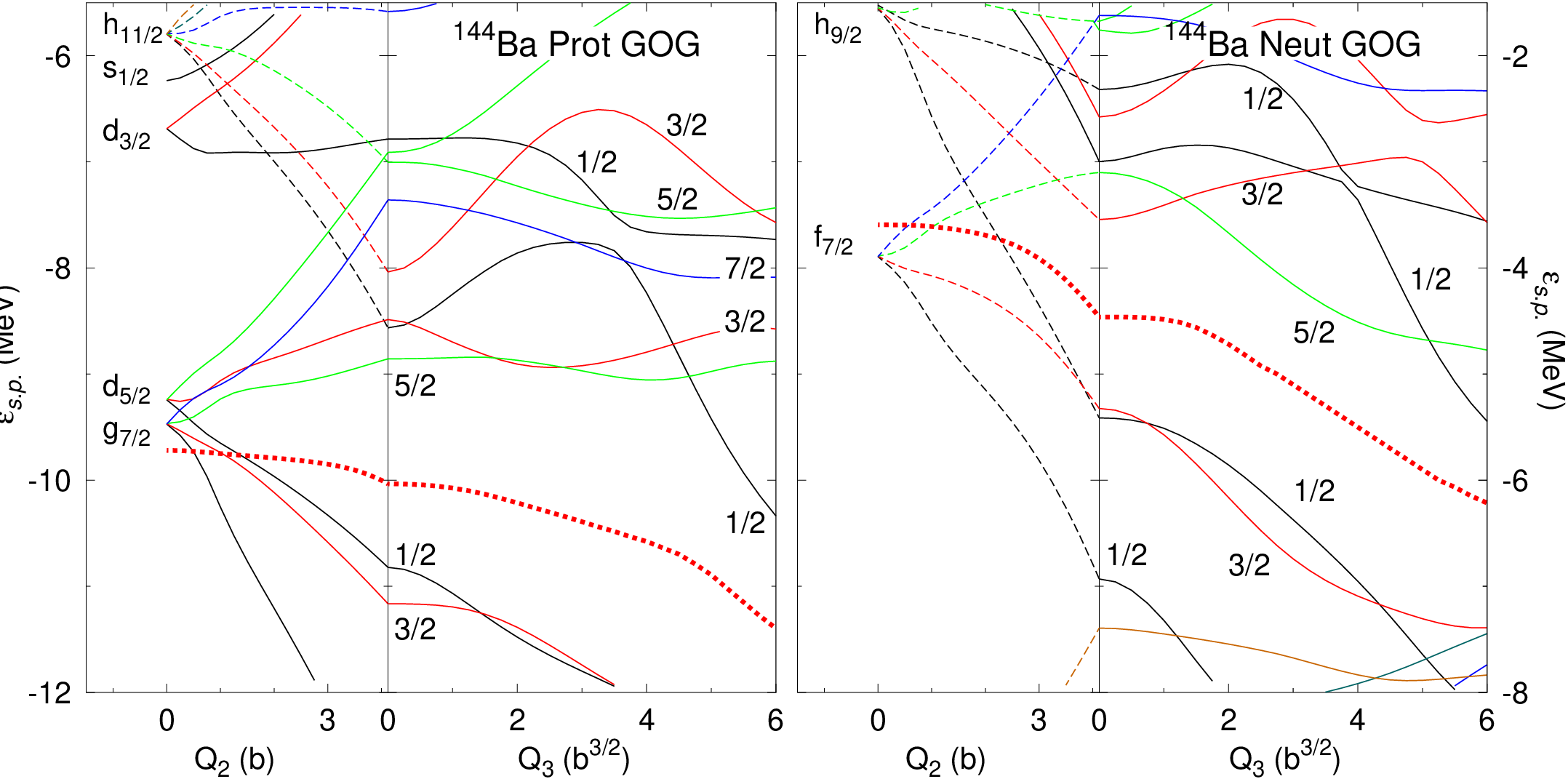}

\includegraphics[width=1.5\columnwidth]{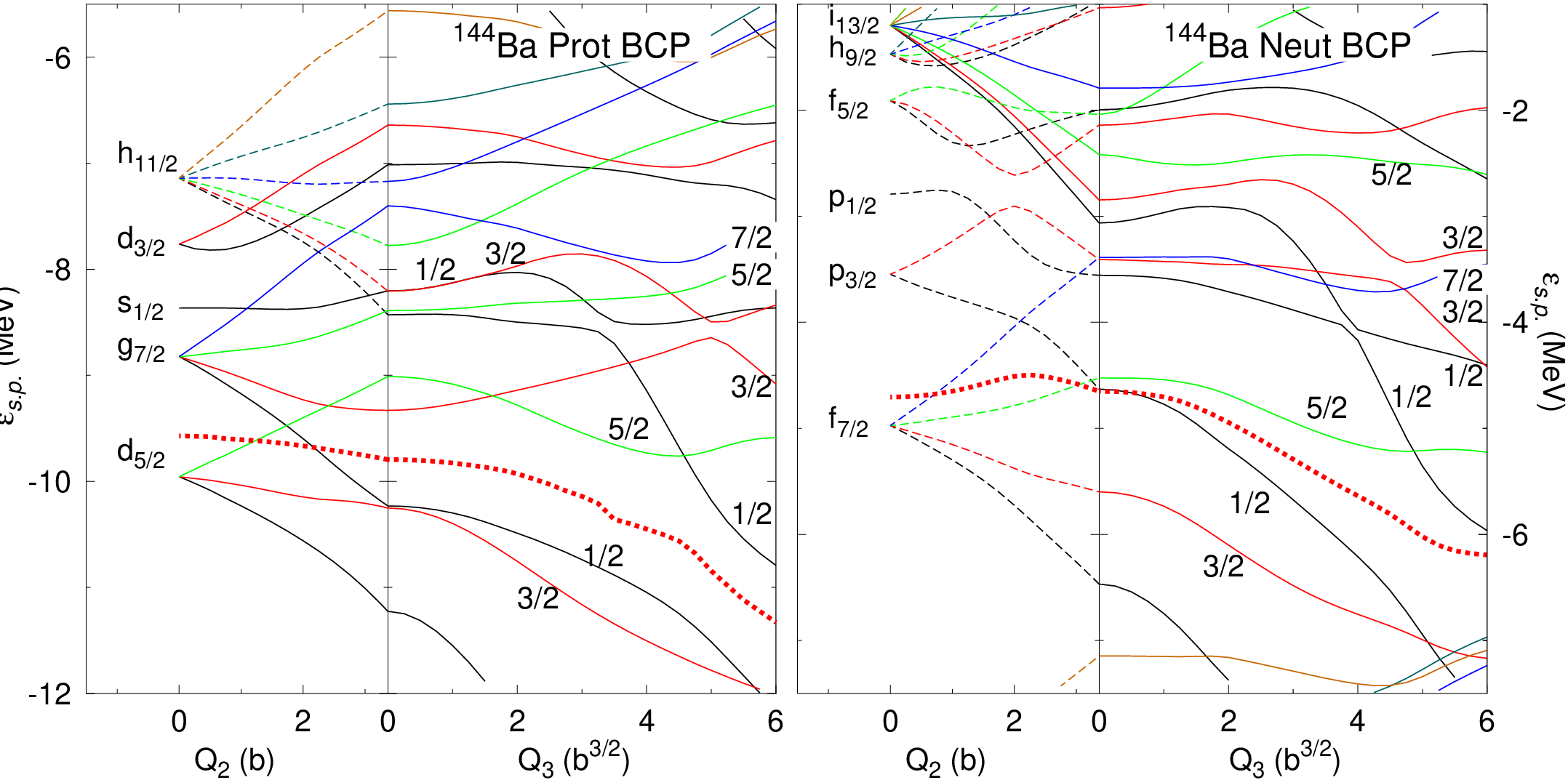}

\caption{(Color online) Single particle energies (SPE) for the nucleus $^{144}$Ba
and computed with the BCP1 energy density functional (lower panels)
and the Gogny D1S (upper panels) are plotted first as a function of
the quadrupole moment $Q_{2}$ (in b) up to the value corresponding
to the self consistent minimum at $Q_{3}=0$ ($Q_{2}=3.37$ b for
the results obtained with the BCP1 functional and 4.12 b for the ones
with the Gogny D1S force) and from there one as a function of the
octupole moment $Q_{3}$(in b$^{3/2}$) (see the caption of Fig \ref{fig:SPE224Ra}
for further details). \label{fig:SPE144Ba}}

\end{figure*}

In Fig \ref{fig:MFPair_Ba} the particle-particle correlation energies
are plotted as a function of the octupole moment. As in the case of
the radium isotopes, we observe that the general trend of $E_{pp}$
both for protons and neutrons and the two interaction/functional considered
is to decrease for increasing octupole moments in the relevant interval
between $Q_{3}=0$ b$^{3/2}$ and $Q_{3}\approx3.5\textrm{b}^{3/2}$;
a tendency that is also observed at higher values of the octupole
moment in most of the barium isotopes studied. Based on the results
of Fig. \ref{fig:MFPair_Ba} as well as the ones of Fig \ref{fig:PAIRING_Ra}
for the radium isotopes it is possible to say, in a very broad sense,
that the onset of octupole deformation tends to quench pairing correlations
and this quenching is bigger for larger values of the octupole moment.
It is also noticed that in most of the cases the pairing correlation
energies obtained in the BCP1 calculation are smaller than the ones
obtained with the Gogny D1S force.

In Fig. \ref{fig:SPE144Ba} the SPE spectrum of $^{144}$Ba obtained
with the two kinds of interactions/functionals is shown. The meaning
of the different panels is the same as in Fig. \ref{fig:SPE224Ra}
for $^{224}$Ra. Here, the same comments made in the analysis and
discussion of Fig. \ref{fig:SPE224Ra} for $^{224}$Ra and regarding
the Nilsson quantum number contents of the single particle wave
functions are also in order.
Coming back to the plot, we observe in the proton spectra the
presence near the Fermi level of an occupied positive parity $d_{5/2}$
spherical orbital and a nearby empty negative parity $h_{11/2}$ orbital.
As discussed in the introduction, this is a characteristic property
of the nuclear SPE for the octupole deformation to take place. In
the SPE for neutrons we observe negative parity orbitals coming from
the $f_{7/2}$ sub-shell below the Fermi level and positive parity
orbitals coming from the $i_{13/2}$ above, which is again a characteristic
signature of octupole deformation. We also observe that the Fermi
level both for protons and neutrons is located in a region of low
level density in the case of the Gogny force calculations, which is
a required condition for octupole deformation to take place (Jahn-Teller
effect). In the case of the calculations with BCP1 we observe that
the Fermi level of neutrons is located in a region of moderate density 
of levels that could explain the lack of octupole deformed minima in
the BCP1 calculations. 

As in the $^{224}$Ra case, the position of the single particle levels 
cannot be used to assign the quantum numbers of neighboring odd-A nuclei 
in a  fashion similar to the Nilsson diagram case.
\begin{figure*}
\includegraphics[width=0.95\textwidth]{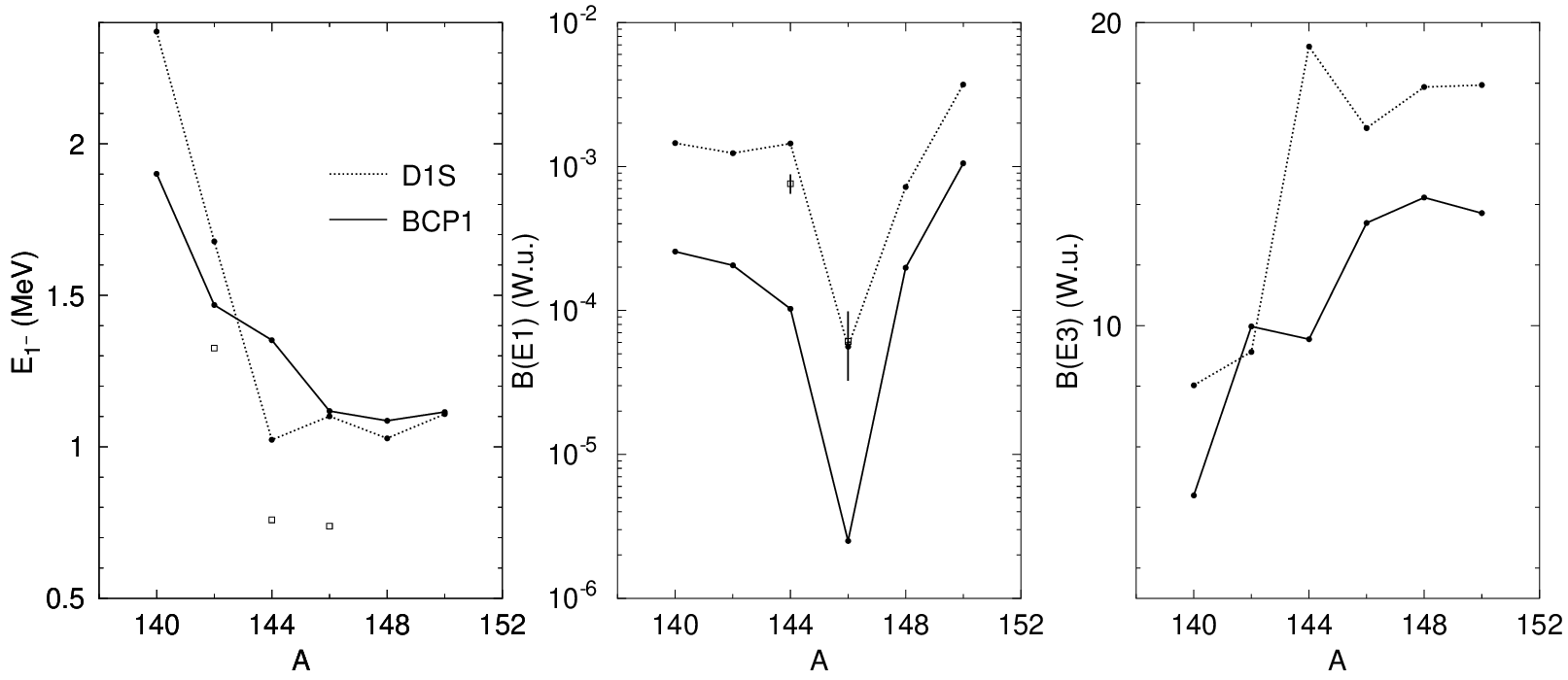}
\caption{Same as Fig. \ref{fig:RES_Ra} but for the even-even barium isotopes
with A in the range from 140 till 150.\label{fig:COLLBa}}

\end{figure*}

In Fig. \ref{fig:COLLBa} the energies of the $1^{-}$ states and
the transition probabilities obtained after solving the CSE are shown
as a function of the mass number of the barium isotopes considered.
In the left hand side panel the excitation energy of the $1^{-}$
state is plotted along with the available experimental data 
\cite{Phillips.96,Urban.97,Shneidman.05,Biswas.05}.
We observe how the isotopic trend is reasonably well reproduced although
the absolute values of the energies are typically a factor of two
larger than the experimental values. The BCP1 excitation energies
are closer to the ones from Gogny D1S than in the radium isotopes
calculations. As discussed previously, the values of the energies
strongly depend upon the values of the collective mass in the vicinity
of the minimum as well as on the height and width of the barrier separating
the two minima with opposite octupole deformation. Consequently, those
excitation energies depend on the interaction/functional used as well
as the level of detail of the theoretical description (inclusion of
pairing correlations, restoration of symmetries) and therefore the
discrepancy with the experiment is not very relevant. Concerning the
$B(E1)$ values, we observe a dip in $^{146}$Ba which is caused by
the same effect as the dip in $^{224}$Ra, namely the peculiar behavior
of the dipole moment $D_{0}(Q_{3})$ with the octupole moment and
mass number. The BCP1 values for the $B(E1)$ transition probabilities
are systematically smaller than the Gogny D1S ones by almost one order
of magnitude. As in the case of the excitation energies of the $1^{-}$
states, it has to be stressed that the isotopic trend is consistent
in the two sets of calculations and both nicely reproduced the scarce
experimental data \cite{Butler.91,Phillips.96}. Finally, in the
right hand side panel of Fig. \ref{fig:COLLBa} the $B(E3)$ transition
probabilities are plotted. In this case and to our knowledge, no experimental
data is available. The $B(E3)$ values show a not very smooth behavior
that, on the other hand, is inversely correlated with the excitation
energies of the $1^{-}$ states. Therefore the bigger $B(E3)$ values
are obtained for the nuclei with the lower $1^{-}$ excitation energies.

\subsection{Fission valley properties of $^{240}$Pu}

\begin{figure}
\includegraphics[width=0.95\columnwidth]{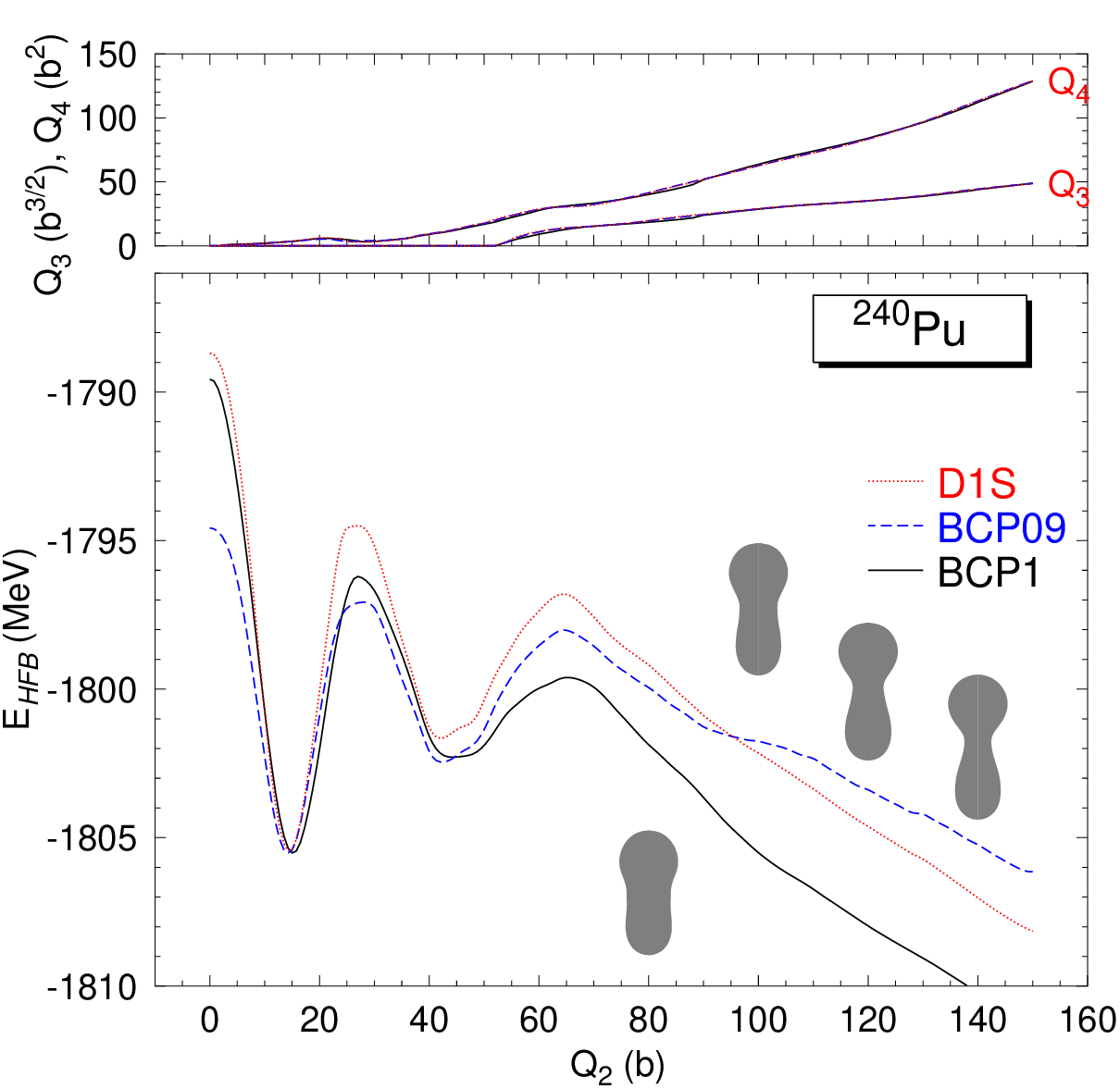}%
\caption{(Color online) In the lower panel the axially symmetric fission path
(mean field energy versus axial quadrupole moment in barns) for the
nucleus $^{240}$Pu and the three interactions/functionals considered
(BCP1 full line, BCP2 dashed line, Gogny D1S dotted line). The shapes
depicted correspond to the half-density contour line of the actual matter
density distribution and are plotted such that their symmetry axis is located at
the corresponding $Q_{20}$ value. In the upper panel the octupole
and hexadecapole moments of the corresponding mean field states. Apparently,
there are only two curves but this is because the results for the
three functionals/interactions lie one on top of each other. \label{fig:240Pu_Fission}}

\end{figure}

In this section the fission properties of $^{240}$Pu, regarding the
octupole contents of the mean field configuration in its way out to
scission, are analyzed for the two BCP functionals \cite{Baldo.08}
and the Gogny D1S force. This isotope has been chosen as a paradigmatic
example of fissioning nucleus that has been thoroughly studied with
the Gogny force \cite{Berger.84}. As it is customary, the theoretical
description of fission is based on the analysis of the potential energy
curve (PEC) obtained by performing constrained mean field calculations
with the quadrupole moment $Q_{2}=\langle Q_{20}\rangle$ as constraining
quantity. Details on the procedures involved can be found in the literature
\cite{Berger.84,Warda.02,Dubrai.08}. The PECs obtained in this
case for the nucleus $^{240}$Pu are shown in the lower panel of Fig.
\ref{fig:240Pu_Fission}. The range of quadrupole moments considered
starts at sphericity ($Q_{2}=0$) and goes up to values corresponding
to very elongated configurations closed to fission ($Q_{2}=150$ b).
At $Q_{2}\approx70$ b the matter distribution starts to resemble
the form of two fission fragments connected by a neck as can be observed
in the shapes given in the figure corresponding to the contour lines
of the matter distribution at half density (see caption for further
details). As the quadrupole moment increases, the distance between
the fragments also increases and the width of the neck decreases.
As a consequence, the energy roughly corresponds to the dominant Coulomb
repulsion between the two incipient fragments. We observe how the
results obtained with BCP and Gogny D1S are fairly similar with the
position of the ground state minimum and the fission isomer lying
at roughly the same $Q_{2}$ values. The potential energy curves obtained
with the BCP1 and BCP2 functionals are very similar and are hardly
distinguishable in Fig. \ref{fig:240Pu_Fission}. The fission isomer
obtained with Gogny D1S lies at an excitation energy almost 2 MeV
higher than the one obtained with the BCP functionals. It is also
observed that the first and second fission barrier heights obtained
with the Gogny D1S force are higher than the ones obtained from the
BCP functionals. This is very likely a consequence of the surface
coefficient in semi-infinite nuclear matter that is higher in Gogny
D1S than in the BCP functionals (see the values quoted in \cite{robledo.08}).
At this point and taking into account the differences in the PECs
it can be concluded that the predictions for the spontaneous fission
half lives obtained with the BCP functionals and Gogny D1S force are
going to be higher for Gogny D1S than for BCP. However, a definitive
answer to this question can not be given until the effect of triaxiality
has been incorporated into the calculations because it is well known
that triaxiality can have a strong impact in the first barrier height.
Also, it has to be kept in mind that the collective mass along the
$Q_{2}$ collective degree of freedom, and entering the WKB formula
used to estimate fission half lives, can be substantially different
when computed with the BCP functionals or the Gogny D1S force. Therefore,
the detailed discussion of the fission half lives obtained with the
BCP functionals is deferred to a more detailed study of fission properties
obtained with this class of functionals. In this paper, devoted to
octupole deformation, it is enough to confirm that the shapes of the
nucleus in its way down to fission are essentially the same irrespective
of the functional/interaction used as can be seen in the upper panel
of Fig. \ref{fig:240Pu_Fission}. In this plot, the octupole and hexadecapole
moments are depicted as a function of $Q_{2}$ and, as can be observed,
the curves for different interactions/functionals are indistinguishable
of each other. This fact implies that the mass distribution close
to scission (as depicted in Fig. \ref{fig:240Pu_Fission} through
the half density contours) is the same irrespective of the interaction/functional
used in the calculation and therefore the predictions of the fission
fragment mass distributions obtained at the mean field level with
Gogny D1S force and BCP functionals should coincide.

\section{Conclusions}

We have explored the octupole degree of freedom in two sets of isotopes
with the newly postulated BCP functionals. The agreement found with
both experiment and the benchmark results obtained in the same framework
with the Gogny D1S interaction, gives us confidence on the good properties
of the BCP functionals concerning odd parity multipole moments. In
addition, the matter distribution of the fissioning nucleus $^{240}$Pu,
which strongly depend upon the response of the system to octupole
perturbations, is found to be essentially the same in the three calculations
performed implying thereby that the BCP functionals and the Gogny
D1S force are equally well suited in that respect. Taking into account
the microscopic origin of the BCP functionals it is comforting and
encouraging to observe its good performance in properties like octupolarity
that belong to the realm of finite nuclei.
\begin{acknowledgments}
Work supported in part by MICINN (FPA2007-66069 and FPA2008-03865-E/IN2P3)
and by the Consolider-Ingenio 2010 program CPAN (CSD2007-00042). X.
V. also acknowledges the support from FIS2008-01661 (Spain and FEDER)
and 2009SGR-1289 (Spain). Support by CompStar, a Research Networking
Programme of the European Science Foundation is also acknowledged.\end{acknowledgments}

\end{document}